\documentclass[10pt,a4paper,twocolumn]{article}
\usepackage[utf8]{inputenc}
\usepackage[english]{babel}
\usepackage{amsmath}
\usepackage{amsfonts}
\usepackage{amssymb}
\usepackage{graphicx}
\usepackage{url}
\usepackage[numbers,compress]{natbib}

\author{Jordi Paillisse$^{*}$, Miquel Ferriol$^{*}$, Eric Garcia$^{*}$, Hamid Latif$^{*}$, Carlos Piris$^{*}$, \\
Albert Lopez$^{*}$, Brenden Kuerbis$^{\dag}$, Alberto Rodriguez-Natal$^{\ddag}$, Vina Ermagan$^{\ddag}$, \\
Fabio Maino$^{\ddag}$ and Albert Cabellos$^{*}$ \\
$^{*}$\textit{\small UPC-BarcelonaTech, Barcelona, Spain} - \textit{\small \{jordip, alopez, acabello\}@ac.upc.edu,} \\
\textit{\small \{miquel.ferriol, eric.garcia.ribera, hamid.latif, carlos.piris\}@est.fib.upc.edu  }\\
$^{\dag}$\textit{\small Georgia Institute of Technology, School of Public Policy, Atlanta, GA, USA } \\
\textit{\small brenden.kuerbis@pubpolicy.gatech.edu} \\
$^{\ddag}$\textit{\small Cisco Systems, San Jose, CA, USA - \{natal,vermagan,fmaino\}@cisco.com}}

\title{IPchain: Securing IP Prefix Allocation and Delegation with Blockchain}

\begin{document}
\maketitle

\begin{abstract}
We present IPchain, a blockchain to store the allocations and delegations of IP addresses, with the aim of easing the deployment of secure interdomain routing systems. Interdomain routing security is of vital importance to the Internet since it prevents unwanted traffic redirections. IPchain makes use of blockchains' properties to provide flexible trust models and simplified management when compared to existing systems. In this paper we argue that Proof of Stake is a suitable consensus algorithm for IPchain due to the unique incentive structure of this use-case. We have implemented and evaluated IPchain's performance and scalability storing around 150k IP prefixes in a 1GB chain.
\end{abstract}
\section{Introduction}

Inter-domain routing security is a pressing issue in today's Internet. In a nutshell, inter-domain routing security encompasses the correct announcement and propagation of IP prefixes across the Autonomous Systems (AS) that conform the Internet. Currently, the protocol that communicates these announcements is BGP (Border Gateway Protocol, RFC 4271). BGP security is typically based on careful configuration via out-of-band mechanisms where network operators tell each other which prefixes to announce. Hence, an accidental misconfiguration or a malicious attacker controlling a BGP router can divert traffic to networks which should not receive it or make ranges of IP addresses unavailable (and effectively bringing Internet services down). This attack is commonly know as prefix hijacking and can be accomplished forging BGP announcements and propagating them to neighboring ASes. We can find several real-life examples of prefix hijacks \cite{Singel2008, Greenberg2014, Dyn2005}.

Given the severity of these attacks, the IETF (Internet Engineering Task Force) has designed a solution to inter-domain routing security by means of the RPKI (Resource Public Key Infrastructure, RFC 6480), a PKI repository to record the legitimate owners of IP prefixes, AS numbers and ROAs (Route Origin Authorization, a certificate to allow an AS to announce an IP prefix).

\begin{figure}[!t]
\centering
\includegraphics[width=0.5\textwidth]{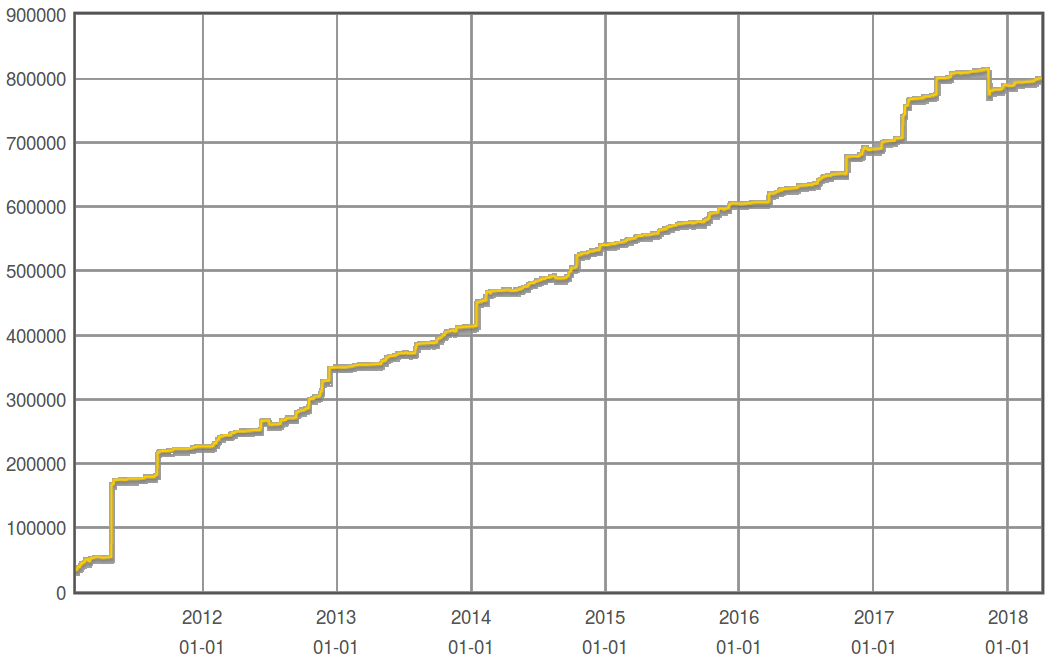}
\caption{Amount of RIPE's IPv4 address space covered by ROAs, in /24 units (source \cite{riperpkistats}). Of the five Registries, RIPE is the one with highest RPKI adoption}
\label{fig:rpkistats}
\end{figure}

Unfortunately, the global deployment of the RPKI is slower than expected with only $\sim$9\% of the total /24 IPv4 address blocks owned by the five Internet Registries being protected by the RPKI (figure \ref{fig:rpkistats}). The reasons of this have been extensively analyzed and discussed in the literature \cite{Wahlisch2015, George2015, Goldberg2014, Liu2016, gilad2016we}, mainly: (i) \emph{Centralized operations:} Certification Authorities (CAs) hold ultimate control of resources in the RPKI. Since IP addresses are a critical asset of RPKI's participants (especially ISPs), they would like to have a higher degree of control over them, but without loosing the security of being certified by a CA, i.e. balanced power between users and CAs \cite{Cooper:2013:RMR:2535771.2535787}. (ii) \emph{Management complexity:} PKIs are cumbersome to manage, e.g. when performing a key rollover. In addition, deploying these extensions is not trivial and requires trained staff \cite{George2015} and financial investment, and (iii) Exposure of business relationships through peering agreements in the RPKI \cite{Wahlisch2015}. In addition, the RPKI faces implementation \cite{Gilad:2017:MCH:3143361.3143363} and transparency \cite{Heilman:2014:CRI:2619239.2626293} challenges.


In the light of this situation, in this paper we propose IPchain: a blockchain to store IP address allocation and delegation data. IPchain leverages some of blockchain's properties to ease the deployment of inter-domain routing security mechanisms. Three of these properties stand out when compared to the RPKI: (i) the ability to create flexible trust models, providing a different balance of power between CAs and downstream users, (ii) simplified management, especially regarding common PKI operations such as key rollover and (iii) auditability: blockchain's append-only ledger can detect possible configuration errors even before a modification\cite{kuerbis2017internet}.

IP addresses share some fundamental characteristics with coins, such as uniqueness or divisibility. Taking advantage of this, IPchain allows its participants to exchange IP prefixes just like in coins are transferred in Bitcoin. This way, an ISP can record in the chain its IP addresses and who can advertise them. When another ISP receives a BGP announcement with these addresses, it can determine if they come from the intended source.

We have built a prototype to demonstrate IPchain's suitability to this scenario, both from the scalability and performance standpoints. The prototype allows allocating and delegating IP prefixes by means of blockchain transactions, and uses a Proof of Stake consensus algorithm to randomly select block signers among all the holders of IP addresses. In our experimental evaluation, we recreated the real-life allocation hierarchy of IP addresses, storing around 50\% of the total Internet Registries' prefixes in a 1 GB chain.




\section{Background: RPKI Architecture}\label{rpki}
IP addresses are allocated to Internet domains following a hierarchical scheme, typically conformed of three tiers (figure \ref{fig:depl_hierarchy}). The Internet Assigned Numbers Authority (IANA\footnote{https://www.iana.org/numbers}), as the top-level regulator of Internet numbers, owns the entire address space. First, IANA allocates large blocks of addresses to the Regional Internet Registries (1). Those, in turn, allocate or delegate blocks to its customers, usually ISPs (2). Finally, ISPs can also assign addresses to their users (3). Equivalently for AS numbers. 

\begin{figure}[!ht]
\centering
\includegraphics[width=0.7\columnwidth, trim={7.5cm 2cm 9cm 2cm},clip]{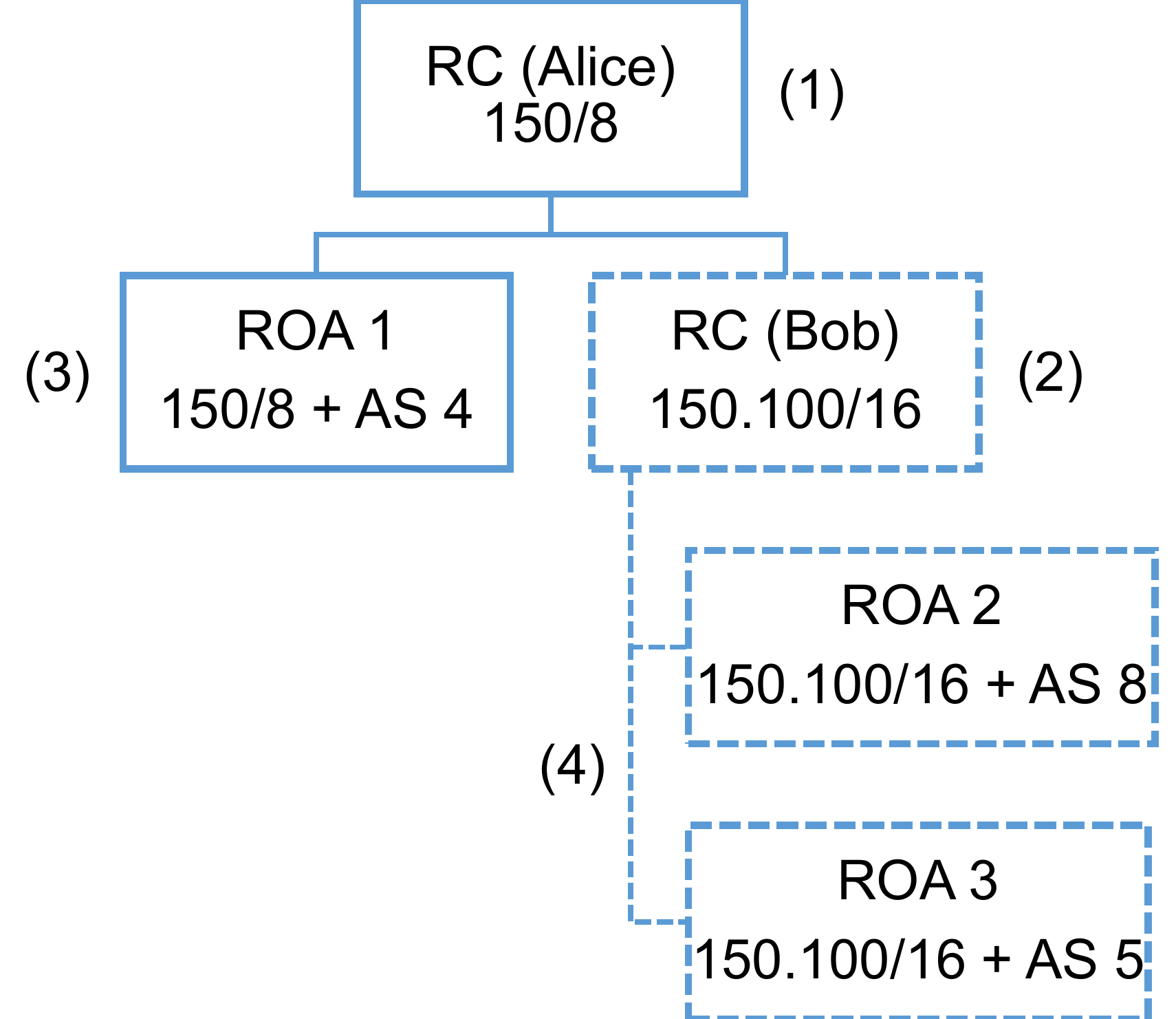}
\caption{Sample RPKI certificate hierarchy.}
\label{fig:rpki}
\end{figure}

The RPKI replicates this structure using digital certificates that allow authenticating the allocation of both IP prefixes and AS numbers (figure \ref{fig:rpki}). A party receives a them via a Resource Certificate (RC), which binds an IP prefix or AS number to a public key (1). It can then further allocate this resource to other parties by means of issuing new RCs (2, dashed line). It can also issue a Route Origin Authorization (ROA), which authorizes an AS number to announce an IP prefix (3). Parties that own RCs can issue one or more ROAs (4).

This way, network operators can download the certificates and use them to verify BGP announcements. If the (IP prefix, AS number) pair in the BGP message does not match the corresponding certificate in the RPKI, the announcement is considered invalid.


\section{Why blockchain?}\label{whybc}
Blockchains have many interesting advantages. In this section we discuss why a blockchain suits the requirements of secure allocation and delegation of IP addresses.

\textbf{\emph{Flexible trust models:}} Some researchers argue that the centralized nature of the RPKI hinders its deployment\cite{Goldberg2014}. This is due to the fact that its users (typically ISPs) have to trust the RPKI CA, which can arbitrarily revoke any downstream certificate \cite{Cooper:2013:RMR:2535771.2535787}. Since IP addresses are an important economical asset for most ISPs, this situation is sometimes judged as inappropriate.

On the contrary, the decentralized trust model of blockchains can mitigate these concerns, since the owner of the public-private key pair retains ultimate control over its resources.

\textbf{\emph{Simplified management:}} the RPKI is cumbersome to manage, for example, users have to choose between two operation modes. Some actions are complex, like key rollover (RFC 6489 is specifically devoted to key rollover in the RPKI\footnote{https://rfc-editor.org/rfc/rfc6489.txt}), because it requires re-signing all downstream certificates starting from the one being replaced. On the contrary, a key rollover in a blockchain can be easily performed transferring a coin/asset to a new address (keypair).  Other operations, such as the revocation of a transaction, do not require a dedicated sub-system (Certificate Revocation Lists (CRLs) in a PKI), but only adding a new transaction. 

\textbf{\emph{Privacy:}} Blockchain transactions are not linked to the user's identity, just to a public key. It is worth noting that the RPKI also offers privacy, because its certificates do not contain identity information.

\textbf{\emph{Consistent vision of the state}}: Exactly like in Bitcoin, in the RPKI we need to keep track of the owner of each IP prefix (coins), e.g. to avoid the transfer of the same prefix to two different users (double-spending). In other words, we need to maintain a global vision of the state. Achieving this is easier in a blockchain when compared to the RPKI: the latter has to update state via specific protocols\footnote{https://rfc-editor.org/rfc/rfc8181.txt}, processing of CRLs and manifests, etc, while in a blockchain these mechanisms directly arise from its transactional nature. 

\textbf{\emph{Auditability:}} Given the permanent nature of blockchain records, it is possible to determine if an object (e.g. a ROA) utilizing a particular resource (e.g. an IP address) has been made obsolete by a new object. In addition, a permanent ledger avoids situations where deleting an object inadvertently impacts other operators \cite{kuerbis2017internet}. While this can be engineered in a PKI\footnote{https://datatracker.ietf.org/wg/trans/about/}, blockchains have this feature built-in.



\section{Which consensus algorithm?}
Consensus algorithms are probably the most important building block of a blockchain. This section details the motivations when deciding which one to use for the use-case of securing IP address allocation and delegation.

\subsection{Proof of Work}

In Proof of Work (PoW), nodes in the blockchain have to solve a complex mathematical problem to add a block, thus requiring some computational effort. The definitive chain is the one with most computing power spent to create. 

Despite its widespread usage, PoW is not suitable for our use case. In financial environments, PoW suits quite well because it is easy to find a large number of participants, either to create new transactions (money transfer is a popular application) or to add blocks (thanks to the block reward mechanism). However, in our use case this does not hold: the potential amount of members is lower than in financial use-cases and it is unclear which would be the reward. 

In addition, PoW blockchains (typically) decouple its users from the management of the chain, i.e., we can transfer money in Bitcoin without having to create blocks. This user-miner separation can impact negatively on the chain, since the capability to add new blocks and the security of the chain itself depend on the computing power of the participants (miners), which is not always aligned with their interest in the well-being of the blockchain (users). Depending on the objectives of an attacker, certain attacks can become profitable. Namely, buying a large quantity of hardware to be able to rewrite the blockchain with false data (e.g., incorrect delegations of IP addresses). This is very costly in blockchains with a high amount of participants such as Bitcoin or Ethereum (in the order of millions), but in a blockcahin for IP addresses it becomes feasible (the current number of Autonomous Systems in the Internet is around 60k\footnote{http://www.cidr-report.org/as2.0/}).




\subsection{Proof of Stake for IP prefix allocation and delegation}\label{sec:pos}
In a Proof of Stake (PoS \cite{ethPosFaq}) blockchain, participants with more assets/coins are more likely to add blocks. In essence, PoS algorithms randomly select one of the users to sign the next block. The selection is randomly weighted according to the  number of coins of each participant, thus ensuring that users with more interest (stake) in the chain contribute proportionally more to it.

With this premise, we advocate that PoS is the most convenient consensus algorithm for our scenario. First, in PoS the capability to alter the blockchain remains within it. This aspect is of particular importance in the context of IPchain: users holding a large number of IP addresses are more likely to add blocks. A key insight of this paper (and a uniqueness of our scenario) is that participants have a reduced incentive in tampering the blockchain because they would suffer the consequences: an insecure Internet. Typically entities that hold large blocks of IP address space have their business within the Internet and as such, have clear incentives in the correct operation and security of the Internet (section 4 in the technical annex). 

Second, the risk of takeover is reduced compared to Proof of Work: accumulating a large amount of IP addresses is typically more complex than accumulating computing power. The risk of takeover is also mitigated compared to other PoS-based blockchains.  In a typical PoS blockchain for financial environments, an attacker would buy tokens from the other parties, who receive a monetary compensation to participate in the attack (c.f. section 2 in the technical annex). However, in a blockchain for IP addresses this would mean buying IP addresses from other parties. These parties do not have a clear incentive to sell their blocks of addresses to the attacker since IP addresses are an important economical asset.

Finally, we must also remark the advantages of PoS algorithms: low computational cost and no need of dedicated hardware, which lowers the entry barrier to collaborate in the blockchain. 

\section{Architecture of IPchain}
\begin{figure}[!t]
\centering
\includegraphics[width=0.8\columnwidth, trim={4cm 2cm 13cm 1.75cm},clip]{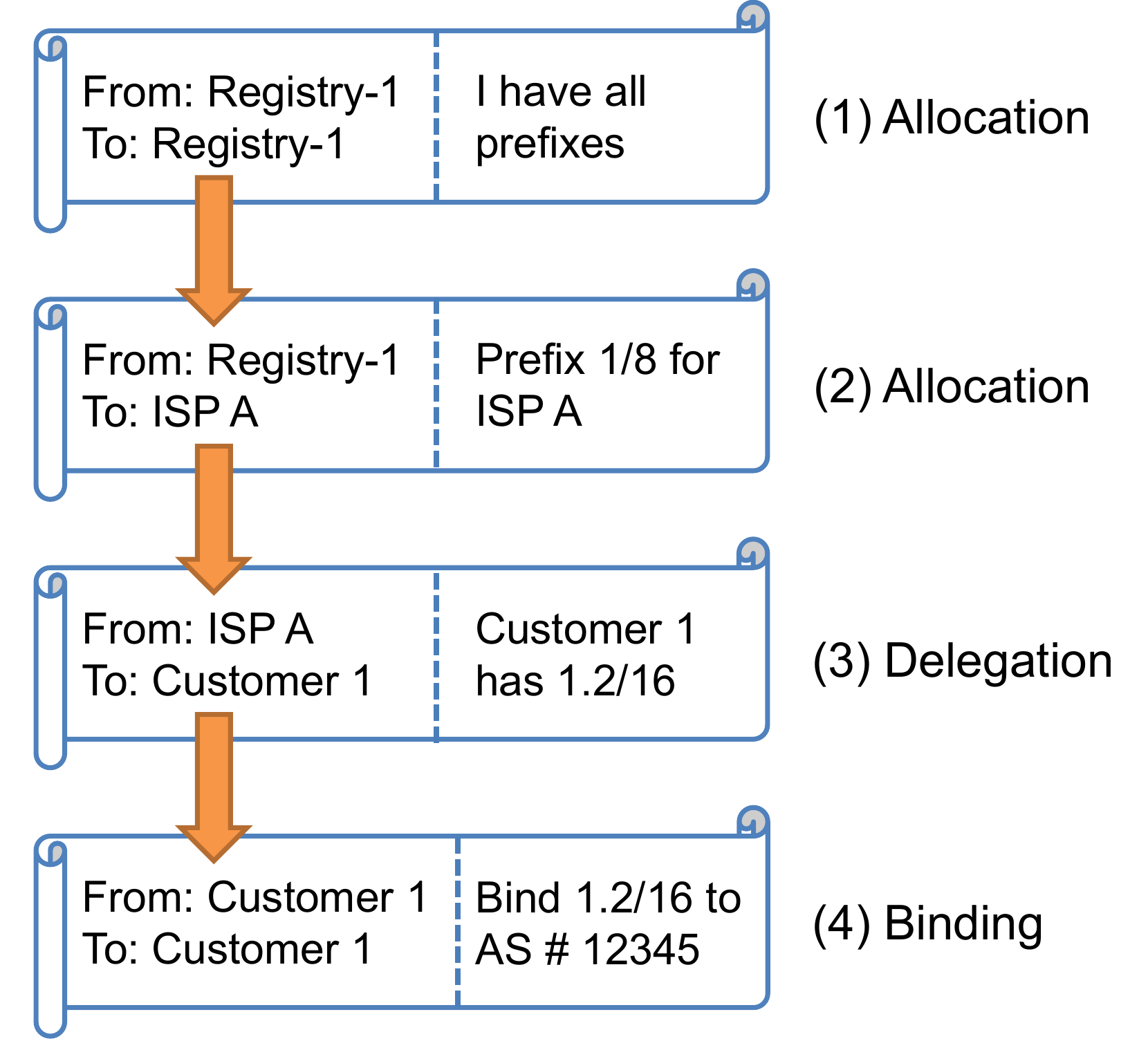}
\caption{Transaction workflow example}
\label{fig:exarch}
\end{figure}

This section describes the architecture of IPchain: a blockchain to store, allocate and delegate blocks of IP addresses.

\subsection{IP prefixes as coins}\label{arch:coins}

IP prefixes share some fundamental characteristics with the coins or assets we find in any blockchain:
\begin{itemize}
\item They are unambiguously allocated to the participants.
\item Can be transferred (delegated) between them.
\item Can be divided up to a certain limit.
\item Cannot be assigned to two participants at the same time.
\end{itemize}

Such similar properties make it possible to design a blockchain to exchange IP prefixes, similarly as if they were crypto-coins. With a series of transactions we can   mimic RPKI's allocation hierarchy in the blockchain and build a consistent view of the legitimate holders of IP prefixes. Figure \ref{fig:exarch} shows an example of this operation: first, Internet Registries (entities that allocate IP addresses) write transactions assigning all the address space to themselves (1). Ideally, this first transaction is encoded in the genesis block. Second, the Registries allocate prefixes to ISPs (2), which in turn allocate them to their customers (3). Finally, customers bind metadata to their prefixes, such as their AS number (4). 

Since the genesis block contains all the prefixes that can be allocated, anyone that downloads the blockchain can validate the chain of allocations and delegations of addresses. If a particular prefix cannot be tracked back to the genesis block, it is invalid.



\subsection{Overview}
\begin{figure}[!t]
\centering
\includegraphics[width=\columnwidth, trim={5.5cm 5cm 9cm 2.5cm},clip]{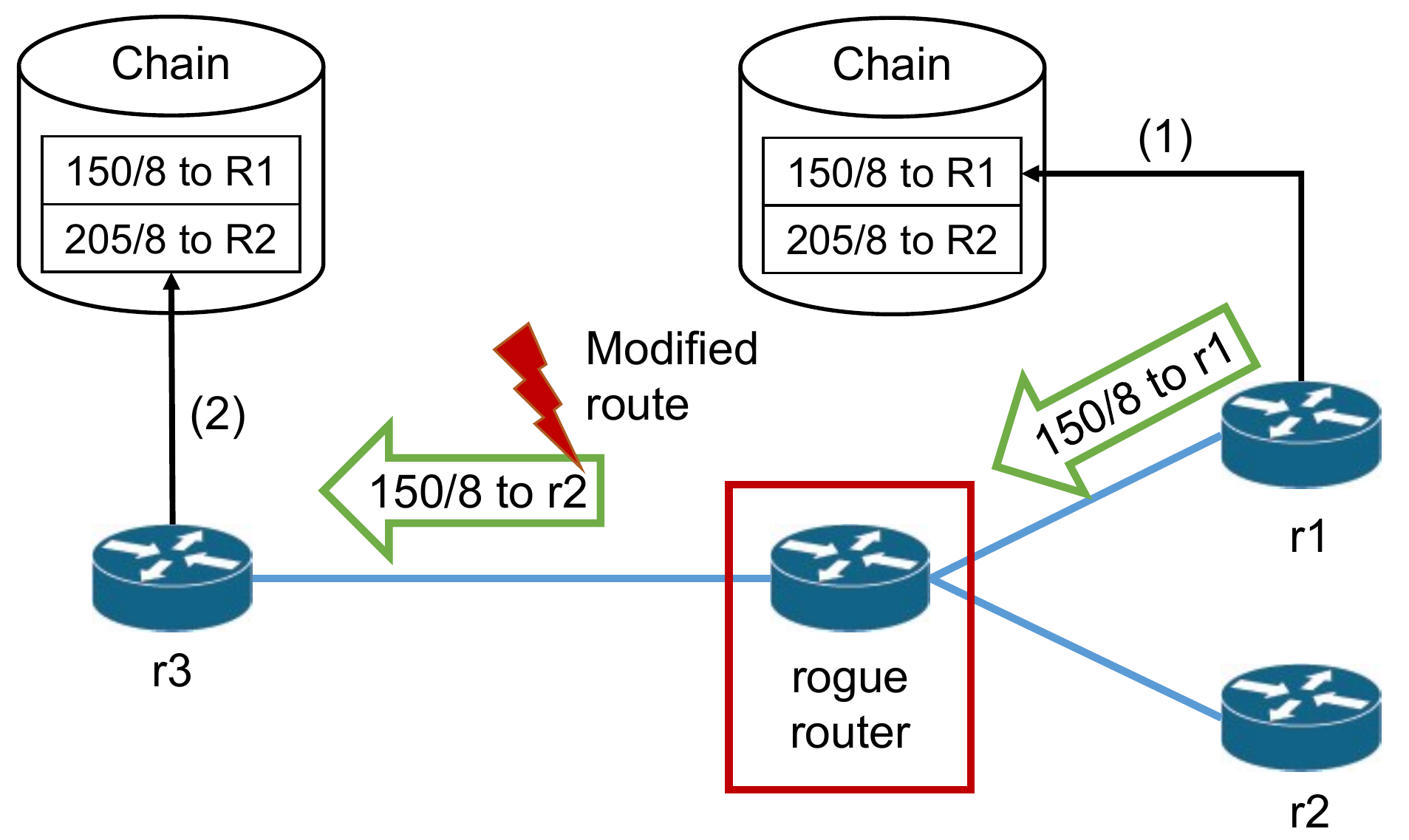}
\caption{Sample usage scenario of IPchain.}
\label{fig:exipcoins}
\end{figure}

Figure \ref{fig:exipcoins} presents a sample of IPchain's workflow. First, router r1 writes in the chain its legitimate prefix (1). Now, consider that the announcement propagates through the network and is modified by the rough router (center). When router r3 receives the announcement of \texttt{150/8 to r2}, it can check in the blockchain (2) if 150/8 is actually originated by router 2. In this case 150/8 should be originated by router r1, so the announcement is deemed invalid.

\subsection{PoS Consensus Algorithm}\label{sec:arch:pos}
As mentioned in section \ref{sec:pos}, PoS is fits the requirements of our use-case. In this scenario, the selection of the next block signer works the following way:
\begin{enumerate}
\item Count the number of addresses owned by each participant
\item Generate a random number
\item Select one of the participants with the random number weighted by their number of addresses
\end{enumerate}
Note that the random number generation is distributed, i.e., all participants generate the same number in order to agree on the same signer. In other words, PoS entails generating a random number in a secure and distributed way. We can find examples in the literature on how to this, such as the Round-Robin Random Number Generator \cite{Baruch2006} or the Shamir Secret Sharing scheme \cite{Shamir:1979:SS:359168.359176}. In addition, more recent algorithms have been proposed specifically designed for blockchains, such as Algorand \cite{Gilad2017} or Ouroboros \cite{ouroboros2017}.

Moreover, in IPchain a reward mechanism (like Bitcoin's block reward) or a punishment for misbehavior (c.f. security deposits in some PoS algorithms) are both undesirable. The former because in this case the incentive is a consistent view of the IP allocation hierarchy and a secure routing infrastructure (and PoS algorithms do not consume high amounts of energy, c.f. section 1 in the technical annex). The latter, because removing IP addresses from a party rarely happens in real-life and generates more severe consequences than not refunding a coin deposit.

\subsection{Supported Operations}\label{sec:arch:ops}
A blockchain for IP addresses has to provide equivalent operations to those performed typically in these process. Similarly to RPKI's logic, we defined the following operations:

\textbf{\emph{Allocate:}} Assign a block of IP prefixes to an entity, allowing it to further allocate or delegate it to other entities.

\textbf{\emph{Delegate:}} Like \emph{Allocate}, but without the permission to allocate prefixes to other entities.\footnote{A delegated prefix cannot be further allocated.}

\textbf{\emph{Metadata:}} Add additional data to a prefix, eg. AS number authorized to announce the prefix.

\subsection{Deployment}\label{sec:arch:deploy}
\begin{figure}[!ht]
\centering
\includegraphics[scale=0.4, trim={12cm 4cm 12.5cm 2.5cm},clip]{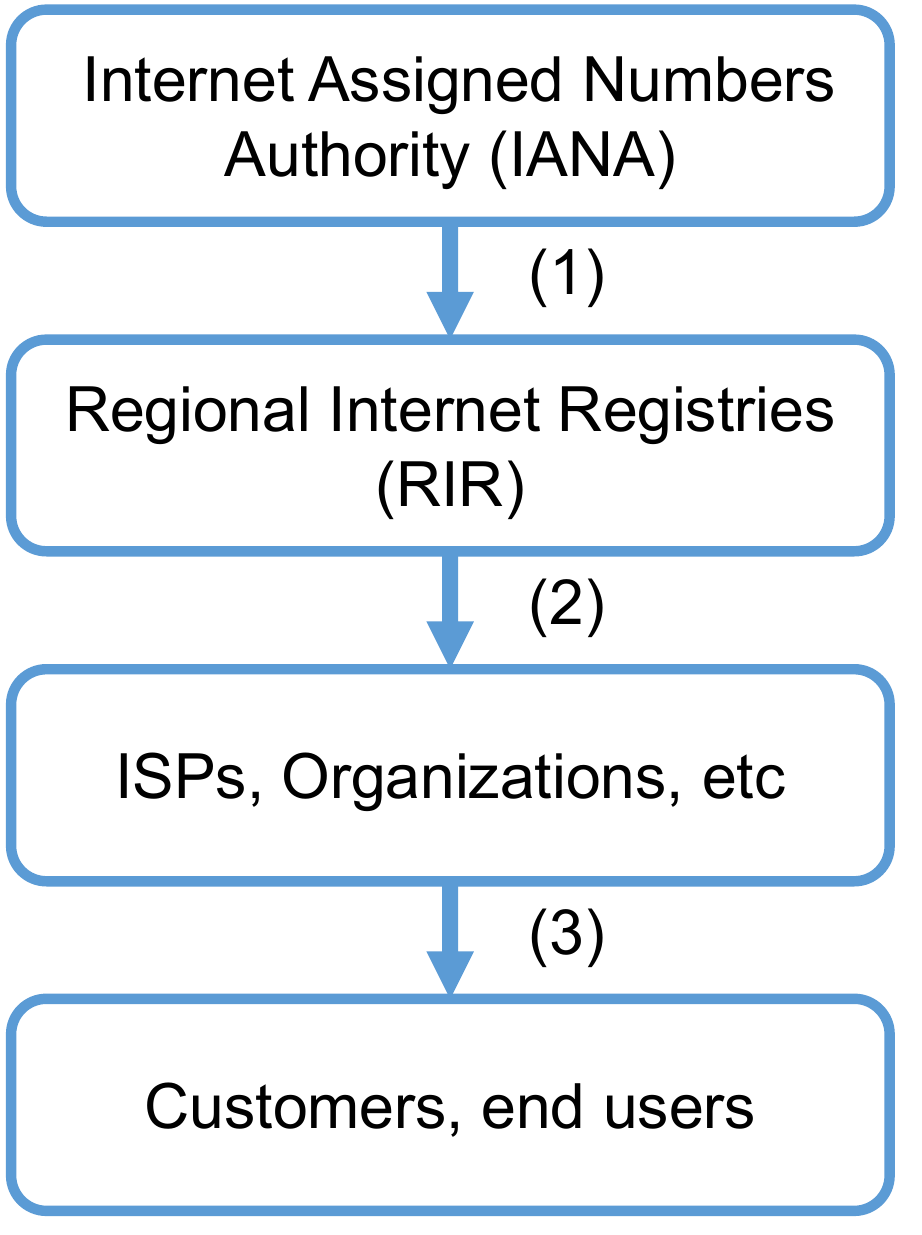}
\caption{IP address allocation hierarchy.}
\label{fig:depl_hierarchy}
\end{figure}

The deployment of our proposed blockchain mimics the current procedure used to allocate IP addresses, which is typically conformed of three tiers (figure \ref{fig:depl_hierarchy}). IANA, as the top-level regulator of Internet numbers, owns the genesis block keys. First, IANA allocates huge blocks of addresses to the Regional Internet Registries (1). Those, in turn, allocate or delegate blocks to its customers, usually ISPs (2). Finally, ISPs can also assign addresses to their users (3).

This way, when bootstrapping the blockchain we only need to trust the code (which has the genesis block embedded). 

\subsection{Flexible Trust: Revocation}\label{sec:revocation}
Due to the irreversible nature of blockchain transactions, once a block of IP addresses has been allocated to an entity it is not possible to modify or remove it, as opposed to CRLs. However, due to operational issues (compromised or lost keys, human mistake, holder misbehavior, etc) it is critical to provide a way to recover a block of addresses. Moreover, since IP addresses are a finite public good they cannot be lost, in contrast to crypto-currencies: the loss of a coin only causes damage to its owner, not the entire community. 

If we consider that a blockchain can enforce any rules its participants agree upon, we can devise some mechanisms that maintain blockchain's inherent decentralization but leave the door open to handle these exceptional situations. A potential approach (but not the only one) is the following: in case of dispute between a RIR and a customer, a third party (e.g. IANA) issues a revocation transaction reassigning the resource.

Despite these remarks, the revocation procedure must be discussed among the community to achieve consensus between the relevant players (IANA, RIRs, ISPs, institutions, etc), taking into account that behind any revocation mechanism there is a fundamental tradeoff between trusting an upstream provider of the addresses (traditional, centralized PKI) and retaining full control of the block of addresses (e.g. Bitcoin-like totally decentralized systems).

\subsection{Other considerations}

\textbf{\emph{Rekeying:}} Rekeying is a common operation in the RPKI and involves re-signing all downstream certificates with the new key. On the contrary, this operation is greatly simplified in blockchain: we only have to add a new transaction re-allocating the IP prefix to a new keypair controlled by ourselves. In addition, since transactions are independent from each other, we can perform rekeying operations individually without affecting other users.

\textbf{\emph{Privacy:}} Since IP addresses are linked to their owners' public key, it is not possible to identify the holder only with the data in the blockchain. In that sense, blockchain offers a similar degree of privacy to the RPKI.

\textbf{\emph{IPv6 support:}} Since IP version 4 and 6 prefixes are currently being used, IPchain needs to support both. However, this is not trivial because there are more IPv6 addresses (128 bits) than IPv4 (32 bits). In a PoS blockchain, randomly selecting from both pools of addresses would create an imbalance of power between and v6 and v4 owners (the first would create much more blocks than the latter).

Taking this into account, we never mix v4 and v6 addresses in IPchain. We create alternatively blocks of v4 or v6 transactions: even blocks contain v4 transactions and are signed by the owner of a v4 prefix. The same for odd blocks and v6 transactions. This solution does not require two different blockchains and isolates v4 and v6 stake.

Finally, it should also be noted that, due to the huge size of v6 address space, large parts of it remain unallocated and still owned by IANA (less than 0.5\% of v6 address space has been allocated to the Regional Internet Registries). This space should be ignored (not counted) to avoid IANA signing nearly all v6 blocks and thus, preventing an IANA monopoly.


\section{Implementation}
We have built an open-source prototype and made it publicly available\footnote{https://github.com/OpenOverlayRouter/blockchain-mapping-system}. We did not to fork an existing blockchain implementation since they do not fit our needs, particularly regarding the consensus algorithm.

\begin{figure}[!t]
\centering
\includegraphics[width=\columnwidth, trim={1cm 3.5cm 1cm 0.5cm},clip]{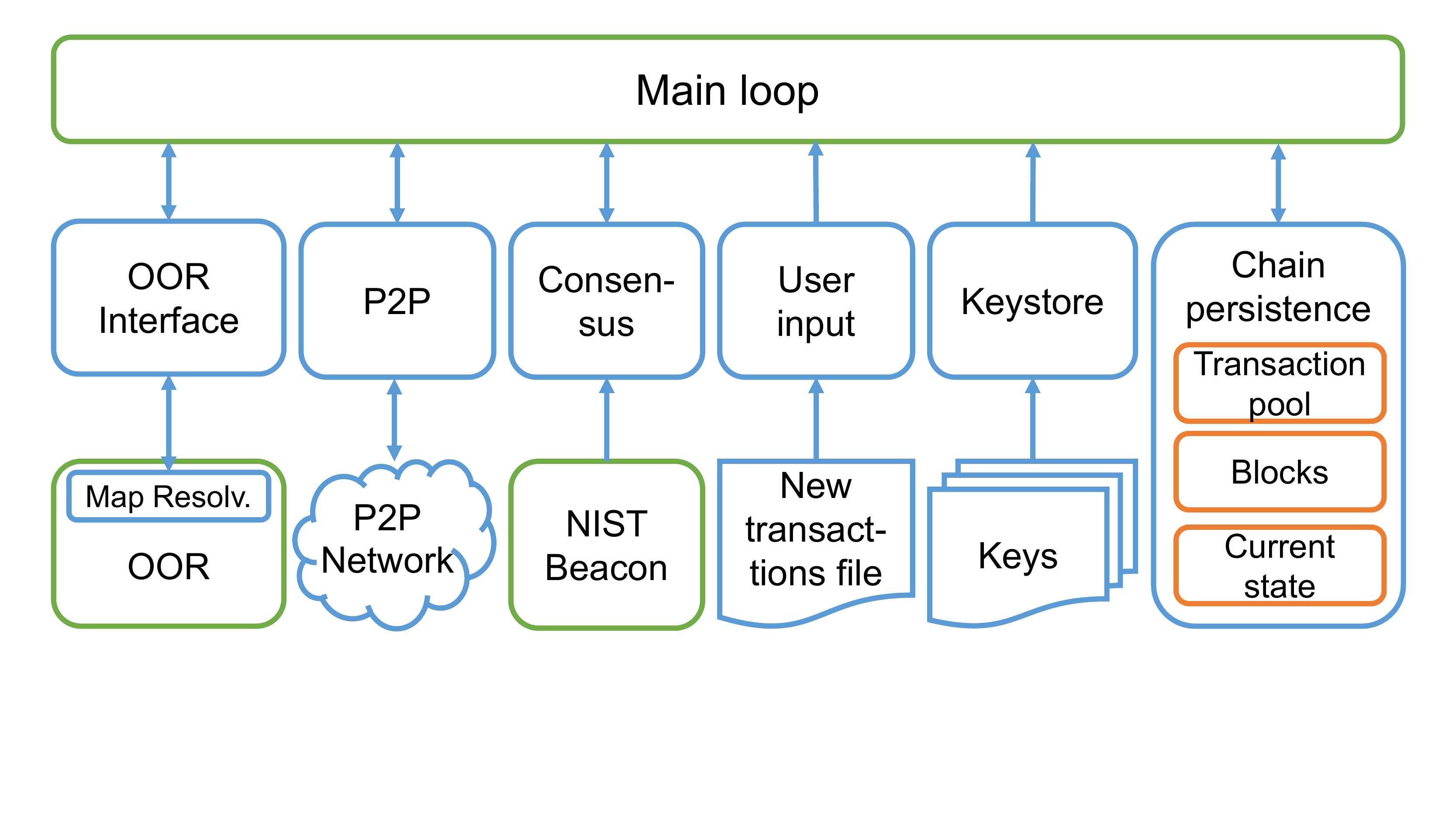}
\caption{IPchain prototype architecture.}
\label{fig:proto-arch}
\end{figure}

The IPchain prototype is written in Python (figure \ref{fig:proto-arch}) and performs all the typical blockchain operations. The transaction structure and validation logic implements the operations defined in section \ref{sec:arch:ops} for both IPv4 and IPv6 addresses. In order to ease user interaction, the prototype reads new transactions from a file, signs and sends them to the network, and provides an integrated keystore to encrypt user's keys.

The prototype has also an interface to communicate with OpenOverlayRouter (OOR \cite{rodriguez2017programmable}). OOR is an open-source software router that implements the Locator/ID Separation Protocol (LISP, RFC 6830). In short, OOR deploys programmable overlay networks to dynamically tunnel traffic through the underlay network. With this interface, OOR can retrieve from IPchain metadata associated with the IP addresses, required to answer some LISP messages.




\textbf{\emph{Data Structures:}} IPchain builds on Ethereum's account system, which maps pairs of blockchain addresses with the associated IP addresses. Transactions are encoded as modifications to accounts. We chose this model instead of Bitcoin's UTXO because it requires less storage and data access is easier. We modified the PyEthereum\footnote{https://github.com/ethereum/pyethereum} Trie, DB, Utils and Transactions classes to fit our needs, and capped the block size at 2 MB.

\textbf{\emph{PoS Consensus Algorithm:}} For simplicity, we use NIST's random beacon\footnote{https://www.nist.gov/programs-projects/nist-randomness-beacon} to select the signer of the next block. This beacon changes every 60 seconds, so we set this interval as our block time. This small blocktime forced us to maintain synchronism among the nodes in the network by means of the Network Time Protocol (NTP). A timeout mechanism selects a new signer when the original is disconnected. Obviously, this cannot be the definitive solution in a production system. 



\textbf{\emph{Peer-to-Peer Network:}} The P2P module implements all communication functions in a broadcast-all fashion, leveraging Pyhton's Twisted\footnote{https://twistedmatrix.com/trac/} library for network communication. Since it does not connect all the nodes between themselves, a Distributed Hash Table\footnote{https://github.com/cliftonm/kademlia-1} keeps track of the last block number, so that if a node misses some blocks it can request them.





\section{Experimental evaluation}

\subsection{Scenario}
The objective of our experiment was to store all the IP prefixes allocated by the five RIRs in the blockchain, so we could evaluate IPchain's suitability and performance in a real-life scenario. To this end, we recreated the aforementioned three-tier hierarchy (section \ref{sec:arch:deploy}) in a cloud service. We deployed 9 virtual machines, each in a different location around the world: North California, North Virginia, S\~ao Paulo, Ireland, Barcelona, Frankfurt, Mumbai, Tokyo and Sydney. One of them was a bootstrap node (that acted as IANA) and the remaining 8 nodes represented the RIRs. We encoded in the genesis block the entire v4 and v6 address space (splitting them in the same blocks specified in IANA's registries of v4\footnote{https://www.iana.org/assignments/ipv4-address-space/ipv4-address-space.xhtml} and v6\footnote{https://www.iana.org/assignments/ipv6-address-space/ipv6-address-space.xhtml} address space), and stored all their keys in the bootstrap node. We then generated a distinct set of transactions for each node to simulate the allocations of IANA and the registries.

\subsection{Methodology}\label{sec:sec:methodology}
We generated three levels of transactions to reproduce the process in figure \ref{fig:depl_hierarchy}. For the first and second levels, we uniformly created smaller prefixes from the ones in the genesis block. The prefixes for the third level were extracted directly from the publicly available lists containing the Registries' address allocations\footnote{https://www.nro.net/statistics/}, so we could record real prefixes in the chain. In all three steps, we allocated the prefixes uniformly among all RIR nodes. In total, we generated around 200k transactions (70k in levels 1 and 2, 130k in level 3), far from the 260k in the RIR files. 



\begin{figure*}[!ht]
\centering
\includegraphics[width=\textwidth, trim={4.5cm 0cm 4cm 0cm},clip]{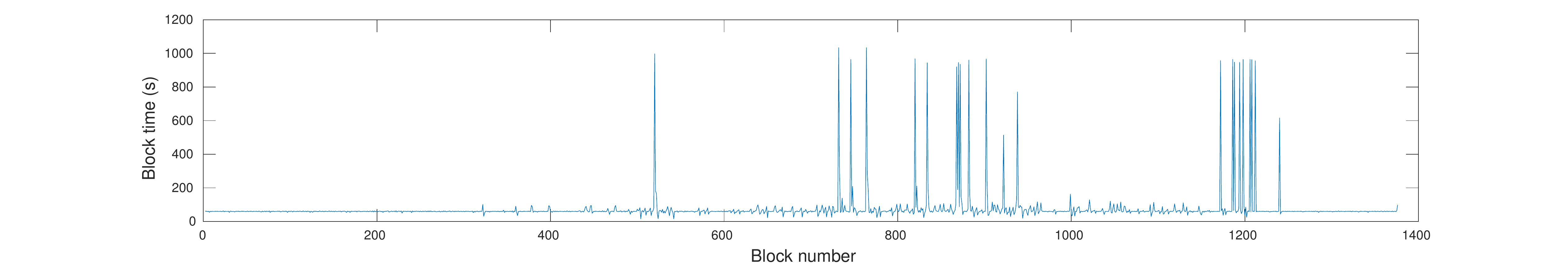}
\caption{Time between each block.}
\label{fig:blocktime}
\end{figure*}
\subsection{Results}

During our test we measured several blockchain metrics, aimed at characterizing the performance and scalability of IPchain. We focus our analysis on four key metrics: (i) throughput, (ii) delay to add a transaction, (iii) bootstrap time, and (iv) chain size. The test run presented here lasted approximately 30 hours with 1376 blocks created and 160k transactions successfully processed. During the experiment some nodes crashed due to communication issues with the NIST beacon. Our prototype recovered automatically (as one would expect from a distributed system). 


\textbf{\emph{Blocktime:}} Figure \ref{fig:blocktime} presents the time between each consecutive block. We can see that in nearly all cases it is around the configured interval of 60 seconds, thus confirming the correct operation of our PoS consensus algorithm. The spikes correspond to the aforementioned node crashes. In that situation, after 900 seconds, a new node was automatically selected by the consensus algorithm to sign that block. The height of the spikes corresponds with these 900 seconds. 

\textbf{\emph{Transaction delay:}} The average delay to add a transaction is usually around 60 seconds, increasing in the same blocks than in figure \ref{fig:blocktime}, due to the delay to create a new block. In our context, this delay is completely acceptable, since this operations in real-life take much more time to process. We infer that we could not overload the system (the minimum time to add a transaction oscillates between 0 and 120 seconds, depending if in this moment a v4 or v6 block is created), otherwise the average delay would be higher. Finding the saturation point remains as future work.



\begin{figure*}[!ht]
\centering
\includegraphics[width=\textwidth, trim={4.5cm 0cm 4cm 0cm},clip]{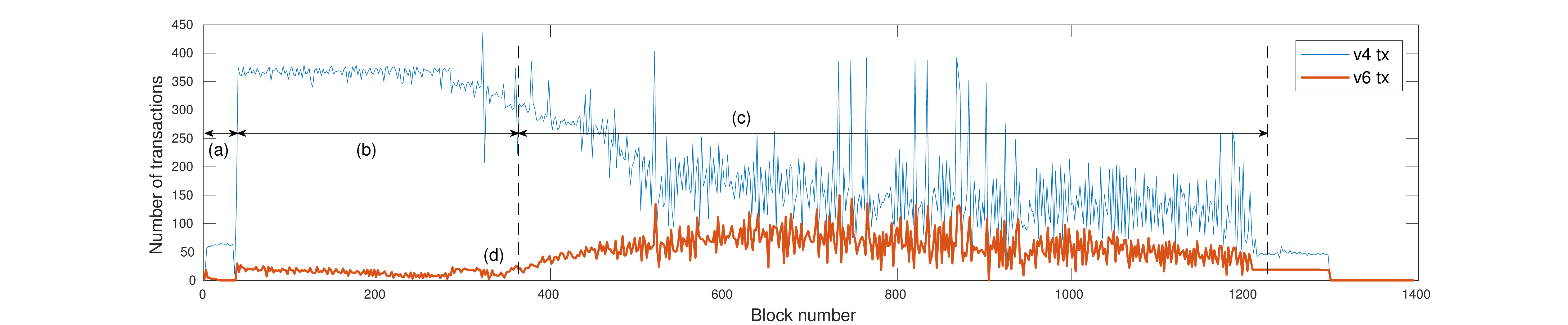}
\caption{Number of transactions in each block. We can see the consequence of the non-uniform distribution of v4 and v6 transactions between blocks 300 and 400 (d, where we start to add RIR v6 transactions): the number of v4 transactions reduces.}
\label{fig:numtx}
\end{figure*}
\textbf{\emph{Throughput:}} Figure \ref{fig:numtx} plots the number of transactions per block, separated for v4 and v6 transactions. The graph outlines the three distinct phases of the experiment detailed in section \ref{sec:sec:methodology}: level 1 (the initial allocation from the bootstrap node to the RIR nodes, a), level 2 (the allocation of prefixes between the RIRs, b), and level 3 (the final allocation of RIR prefixes, c). In the latter, we appreciate a high variability in the number of transactions in a block, roughly spanning from 100 to 250 transactions for v4 blocks and from 50 to 100 for v6 blocks. This is due to several reasons: (i) the non-uniform distribution of v4 and v6 transactions in the input files: the proportion of v4 and v6 transactions is not constant for the entire file. Since nodes process transactions at a fixed speed, regardless if they are v4 or v6, in a given period of time a node may not issue a constant number of v4 or v6 transactions. (ii) the crash of a node: after a restart, nodes wait 35 minutes before starting to add transactions, and (iii) nodes had different number of transactions, so they stopped adding transactions at different moments in time. Again, the spikes in the number of transactions correspond to the aforementioned delayed blocks, due to the pending transactions accumulated during these intervals. In addition, some of these transactions may have been considered invalid because they were referencing transactions not yet included in the chain, further increasing this variability.

The maximum number of transactions per block revolves around 370 (section (b) in figure \ref{fig:numtx}), so we can estimate IPchain's throughput to be approximately 6 transactions per second, in the same order of magnitude of Bitcoin. 





\subsubsection{Bootstrap test}

\begin{figure}[!ht]
\centering
\includegraphics[width=\columnwidth, trim={2cm 0.2cm 0.25cm 1.5cm},clip]{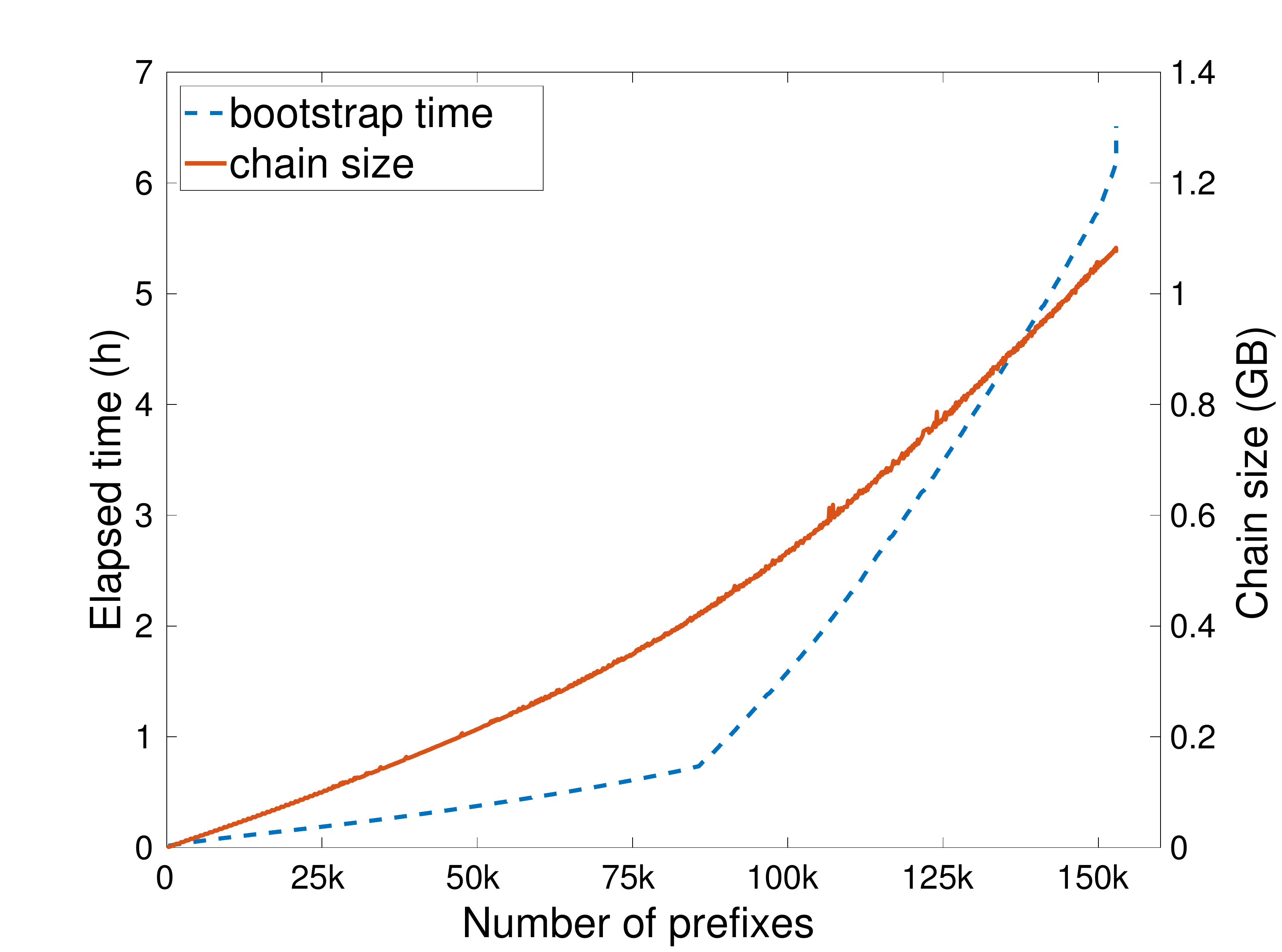}
\caption{Bootstrap test: startup time and total chain size.}
\label{fig:bootstrap}
\end{figure}

The bootstrap test is of special interest, because it quantifies IPchain's overall cost in terms of time and compute resources. After storing all the allocations in the chain, we added a new node to the network and measured how long it took to verify the entire chain (figure \ref{fig:bootstrap}). The time to download the blocks is negligible compared to the validation time. Using a VM with one associated virtual CPU (Intel Xeon @ 3.3 GHz) and 2 GB RAM, the bootstrap time was 7 hours. 

We also recorded the total chain size as we added more prefixes (figure \ref{fig:bootstrap}), reaching around 1 GB for 150k prefixes. In this case, the size grows in a linear fashion with the number of prefixes. Since it is in the order of GB, we can conclude that the chain scales well for our use-case.

\section{Related work}
\textbf{\emph{Blockchain Applications for Networking:}} Several blockchain applications oriented for networks have been proposed \cite{Bozic2016}, such as BGP announcements \cite{Hari2016}, the Internet of Things \cite{Christidis2016}, Information Centric Networking \cite{Fotiou2016} or distributed access control \cite{Maesa2017}. However, the largest body of work focuses on providing naming applications with similar functions to the DNS: Namecoin\footnote{https://namecoin.org/}, Blockstack \cite{ali2016blockstack}, Ethereum Name System\footnote{https://ens.domains/}, etc. To the best of our knowledge, IPchain is the first blockchain specifically tailored for the allocation and delegation of IP addresses.

\textbf{\emph{PoS Consensus Algorithms:}} Some interesting PoS algorithms are already in production systems, such as NEM's Proof of Importance \cite{nemPoi2018}, a modified version of PoS that takes into account not only the stake but also the transaction graph topology and frequency. However, in IPchain we cannot rely on the frequency of transactions like NEM because our scenario is not as dynamic as financial systems. 

Another promising PoS algorithm is Ethereum's Casper \cite{buterin2017casper}, still in development. However, Casper uses a punishment mechanism (remove part of the security deposit) that, as we described earlier in sec \ref{sec:arch:pos} does not make sense here.


\textbf{\emph{Social Network Consensus Algorithms:}} In such algorithms, like Stellar \cite{mazieres2015stellar}, each user has a set of trusted nodes, and only accepts transactions if they have been validated by these nodes. Although they present an interesting alternative, we consider them less suitable for IPchain than Proof of Stake, because they typically require some kind of certificate-based identification system for the nodes, thus losing one of IPchain's fundamental advantages. 




\section{Conclusion}
In this paper we have introduced IPchain, a Proof of Stake blockchain to store the allocation and delegations of IP addresses. We have discussed its advantages over existing systems, such as simplified management or flexible trust models, and the benefits of a Proof of Stake consensus algorithm in this particular context. Finally, we have presented performance data of our implementation when storing RIR data.

\section*{Acknowledgments}

The authors would like to thank Jordi Herrera-Joancomarti, Andreu Rodriguez-Donaire and Jordi Baylina for their helpful discussions about Bitcoin and blockchain technology, as well as Leo Vegoda for his insights regarding revocation mechanisms. This work has been partially supported by the Spanish Ministry of Economy and Competitiveness under contract TEC2017-90034-C2-1-R (ALLIANCE project) that receives funding from FEDER, and by the Catalan Institution for Research and Advanced Studies (ICREA).

\bibliographystyle{unsrtnat}
\bibliography{bcbiblio}

\begin{thebibliography}{30}
\providecommand{\natexlab}[1]{#1}
\providecommand{\url}[1]{\texttt{#1}}
\expandafter\ifx\csname urlstyle\endcsname\relax
  \providecommand{\doi}[1]{doi: #1}\else
  \providecommand{\doi}{doi: \begingroup \urlstyle{rm}\Url}\fi

\bibitem[Singel(2008)]{Singel2008}
Ryan Singel.
\newblock Pakistan's accidental youtube re-routing exposes trust flaw in net,
  February 2008.
\newblock URL \url{https://www.wired.com/2008/02/pakistans-accid/}.

\bibitem[Greenberg(2014)]{Greenberg2014}
Adny Greenberg.
\newblock Hacker redirects traffic from 19 internet providers to steal
  bitcoins, July 2014.
\newblock URL \url{https://www.wired.com/2014/08/isp-bitcoin-theft/}.

\bibitem[Blogs(2005)]{Dyn2005}
Dyn~Guest Blogs.
\newblock Internet-wide catastrophe - last year, December 2005.
\newblock URL \url{https://dyn.com/blog/internetwide-nearcatastrophela/}.

\bibitem[W{\"{a}}hlisch et~al.(2015)W{\"{a}}hlisch, Schmidt, Schmidt, Maennel,
  Uhlig, and Tyson]{Wahlisch2015}
Matthias W{\"{a}}hlisch, Robert Schmidt, Thomas~C Schmidt, Olaf Maennel, Steve
  Uhlig, and Gareth Tyson.
\newblock {RiPKI: The Tragic Story of RPKI Deployment in the Web Ecosystem}.
\newblock \emph{Proceedings of the 14th ACM Workshop on Hot Topics in Networks
  - HotNets-XIV}, pages 1--7, 2015.
\newblock \doi{10.1145/2834050.2834102}.
\newblock URL \url{http://dl.acm.org/citation.cfm?id=2834050.2834102}.

\bibitem[George(2014)]{George2015}
Wes George.
\newblock {Adventures in RPKI (non) deployment}.
\newblock Technical report, NANOG 62, 2014.

\bibitem[Goldberg(2014)]{Goldberg2014}
Sharon Goldberg.
\newblock Why is it taking so long to secure internet routing?
\newblock \emph{Queue}, 12\penalty0 (8):\penalty0 20:20--20:33, August 2014.
\newblock ISSN 1542-7730.
\newblock \doi{10.1145/2668152.2668966}.
\newblock URL \url{http://doi.acm.org/10.1145/2668152.2668966}.

\bibitem[Liu et~al.(2016)Liu, Yan, Geng, Lee, Tseng, and Ku]{Liu2016}
Xiaowei Liu, Zhiwei Yan, Guanggang Geng, Xiaodong Lee, Shian~Shyong Tseng, and
  Ching~Heng Ku.
\newblock {RPKI deployment: Risks and alternative solutions}.
\newblock In \emph{Advances in Intelligent Systems and Computing}, volume 387,
  pages 299--310, 2016.
\newblock ISBN 9783319232034.
\newblock \doi{10.1007/978-3-319-23204-1_30}.

\bibitem[Gilad et~al.(2016)Gilad, Cohen, Herzberg, Schapira, and
  Shulman]{gilad2016we}
Yossi Gilad, Avichai Cohen, Amir Herzberg, Michael Schapira, and Haya Shulman.
\newblock Are we there yet? on rpki's deployment and security.
\newblock \emph{IACR Cryptology ePrint Archive}, 2016:\penalty0 1010, 2016.

\bibitem[Cooper et~al.(2013)Cooper, Heilman, Brogle, Reyzin, and
  Goldberg]{Cooper:2013:RMR:2535771.2535787}
Danny Cooper, Ethan Heilman, Kyle Brogle, Leonid Reyzin, and Sharon Goldberg.
\newblock On the risk of misbehaving rpki authorities.
\newblock In \emph{Proceedings of the Twelfth ACM Workshop on Hot Topics in
  Networks}, HotNets-XII, pages 16:1--16:7, New York, NY, USA, 2013. ACM.
\newblock ISBN 978-1-4503-2596-7.
\newblock \doi{10.1145/2535771.2535787}.
\newblock URL \url{http://doi.acm.org/10.1145/2535771.2535787}.

\bibitem[Gilad et~al.(2017{\natexlab{a}})Gilad, Sagga, and
  Goldberg]{Gilad:2017:MCH:3143361.3143363}
Yossi Gilad, Omar Sagga, and Sharon Goldberg.
\newblock Maxlength considered harmful to the rpki.
\newblock In \emph{Proceedings of the 13th International Conference on Emerging
  Networking EXperiments and Technologies}, CoNEXT '17, pages 101--107, New
  York, NY, USA, 2017{\natexlab{a}}. ACM.
\newblock ISBN 978-1-4503-5422-6.
\newblock \doi{10.1145/3143361.3143363}.

\bibitem[Heilman et~al.(2014)Heilman, Cooper, Reyzin, and
  Goldberg]{Heilman:2014:CRI:2619239.2626293}
Ethan Heilman, Danny Cooper, Leonid Reyzin, and Sharon Goldberg.
\newblock From the consent of the routed: Improving the transparency of the
  rpki.
\newblock In \emph{Proceedings of the 2014 ACM Conference on SIGCOMM}, SIGCOMM
  '14, pages 51--62, New York, NY, USA, 2014. ACM.
\newblock ISBN 978-1-4503-2836-4.
\newblock \doi{10.1145/2619239.2626293}.

\bibitem[rip(2018)]{riperpkistats}
Rpki certification statistics, March 2018.
\newblock URL \url{http://certification-stats.ripe.net/}.

\bibitem[Kuerbis and Mueller(2017)]{kuerbis2017internet}
Brenden Kuerbis and Milton Mueller.
\newblock Internet routing registries, data governance, and security.
\newblock \emph{Journal of Cyber Policy}, 2\penalty0 (1):\penalty0 64--81,
  2017.

\bibitem[Foundation(2018{\natexlab{a}})]{ethPosFaq}
Ethereum Foundation.
\newblock Proof of stake faq, April 2018{\natexlab{a}}.
\newblock URL \url{https://github.com/ethereum/wiki/wiki/Proof-of-Stake-FAQ}.

\bibitem[Awerbuch and Scheideler(2006)]{Baruch2006}
Baruch Awerbuch and Christian Scheideler.
\newblock Robust random number generation for peer-to-peer systems.
\newblock In Mariam Momenzadeh Alexander~A. Shvartsman, editor,
  \emph{Principles of Distributed Systems}, pages 275--289, Berlin, Heidelberg,
  2006. Springer Berlin Heidelberg.
\newblock ISBN 978-3-540-49991-6.

\bibitem[Shamir(1979)]{Shamir:1979:SS:359168.359176}
Adi Shamir.
\newblock How to share a secret.
\newblock \emph{Commun. ACM}, 22\penalty0 (11):\penalty0 612--613, November
  1979.
\newblock ISSN 0001-0782.
\newblock \doi{10.1145/359168.359176}.

\bibitem[Gilad et~al.(2017{\natexlab{b}})Gilad, Hemo, Micali, Vlachos, and
  Zeldovich]{Gilad2017}
Yossi Gilad, Rotem Hemo, Silvio Micali, Georgios Vlachos, and Nickolai
  Zeldovich.
\newblock {Algorand: Scaling byzantine agreements for cryptocurrencies}.
\newblock \emph{Proceedings of the 26th Symposium on Operating Systems
  Principles}, pages 51--68, 2017{\natexlab{b}}.
\newblock \doi{10.1145/3132747.3132757}.

\bibitem[Kiayias et~al.(2017)Kiayias, Russell, David, and
  Oliynykov]{ouroboros2017}
Aggelos Kiayias, Alexander Russell, Bernardo David, and Roman Oliynykov.
\newblock Ouroboros: A provably secure proof-of-stake blockchain protocol.
\newblock In Jonathan Katz and Hovav Shacham, editors, \emph{Advances in
  Cryptology -- CRYPTO 2017}, pages 357--388, Cham, 2017. Springer
  International Publishing.
\newblock ISBN 978-3-319-63688-7.

\bibitem[Rodriguez-Natal et~al.(2017)Rodriguez-Natal, Paillisse, Coras,
  Lopez-Bresco, Jakab, Portoles-Comeras, Natarajan, Ermagan, Meyer, Farinacci,
  et~al.]{rodriguez2017programmable}
Alberto Rodriguez-Natal, Jordi Paillisse, Florin Coras, Albert Lopez-Bresco,
  Lorand Jakab, Marc Portoles-Comeras, Preethi Natarajan, Vina Ermagan, David
  Meyer, Dino Farinacci, et~al.
\newblock Programmable overlays via openoverlayrouter.
\newblock \emph{IEEE Communications Magazine}, 55\penalty0 (6):\penalty0
  32--38, 2017.

\bibitem[Bozic et~al.(2016)Bozic, Pujolle, and Secci]{Bozic2016}
N.~Bozic, G.~Pujolle, and S.~Secci.
\newblock A tutorial on blockchain and applications to secure network
  control-planes.
\newblock In \emph{2016 3rd Smart Cloud Networks Systems (SCNS)}, pages 1--8,
  Dec 2016.
\newblock \doi{10.1109/SCNS.2016.7870552}.

\bibitem[Hari and Lakshman(2016)]{Hari2016}
Adiseshu Hari and T~V Lakshman.
\newblock {The Internet Blockchain : A Distributed , Tamper-Resistant
  Transaction Framework for the Internet}.
\newblock \emph{Fifteenth ACM Workshop on Hot Topics in Networks}, pages
  204--210, 2016.
\newblock \doi{10.1145/3005745.3005771}.

\bibitem[Christidis and Devetsikiotis(2016)]{Christidis2016}
K.~Christidis and M.~Devetsikiotis.
\newblock Blockchains and smart contracts for the internet of things.
\newblock \emph{IEEE Access}, 4:\penalty0 2292--2303, 2016.
\newblock \doi{10.1109/ACCESS.2016.2566339}.

\bibitem[Fotiou and Polyzos(2016)]{Fotiou2016}
N.~Fotiou and G.~C. Polyzos.
\newblock Decentralized name-based security for content distribution using
  blockchains.
\newblock In \emph{2016 IEEE Conference on Computer Communications Workshops
  (INFOCOM WKSHPS)}, pages 415--420, April 2016.
\newblock \doi{10.1109/INFCOMW.2016.7562112}.

\bibitem[Di~Francesco~Maesa et~al.(2017)Di~Francesco~Maesa, Mori, and
  Ricci]{Maesa2017}
Damiano Di~Francesco~Maesa, Paolo Mori, and Laura Ricci.
\newblock Blockchain based access control.
\newblock In Lydia~Y. Chen and Hans~P. Reiser, editors, \emph{Distributed
  Applications and Interoperable Systems}, pages 206--220, Cham, 2017. Springer
  International Publishing.
\newblock ISBN 978-3-319-59665-5.

\bibitem[Ali et~al.(2016)Ali, Nelson, Shea, and Freedman]{ali2016blockstack}
Muneeb Ali, Jude~C Nelson, Ryan Shea, and Michael~J Freedman.
\newblock Blockstack: A global naming and storage system secured by
  blockchains.
\newblock In \emph{USENIX Annual Technical Conference}, pages 181--194, 2016.

\bibitem[Foundation(2018{\natexlab{b}})]{nemPoi2018}
NEM.io Foundation.
\newblock Nem tech reference, April 2018{\natexlab{b}}.
\newblock URL \url{https://nem.io/wp-content/themes/nem/files/NEM_techRef.pdf}.

\bibitem[Buterin and Griffith(2017)]{buterin2017casper}
Vitalik Buterin and Virgil Griffith.
\newblock Casper the friendly finality gadget.
\newblock \emph{arXiv preprint arXiv:1710.09437}, 2017.

\bibitem[Mazieres(2015)]{mazieres2015stellar}
David Mazieres.
\newblock The stellar consensus protocol: A federated model for internet-level
  consensus.
\newblock \emph{Stellar Development Foundation}, 2015.

\bibitem[Buterin(2014)]{ethRangAtt}
Vitalik Buterin.
\newblock Slasher ghost, and other developments in proof of stake, October
  2014.
\newblock URL
  \url{https://blog.ethereum.org/2014/10/03/slasher-ghost-developments-proof-stake/}.

\bibitem[Apostolaki et~al.(2017)Apostolaki, Zohar, and
  Vanbever]{apostolaki2017}
Maria Apostolaki, Aviv Zohar, and Laurent Vanbever.
\newblock Hijacking bitcoin: Routing attacks on cryptocurrencies.
\newblock In \emph{Security and Privacy (SP), 2017 IEEE Symposium on}, pages
  375--392. IEEE, 2017.

\end{thebibliography}

\appendix
\section{Security Analysis} \label{sec:annex}

This technical annex presents a list of the most relevant attacks against our Proof of Stake algorithm, and strategies on how to mitigate them.

\subsection{Nothing at Stake}\label{security:nats}
Nothing at stake \cite{ethPosFaq} is one of the fundamental drawbacks of Proof of Stake and requires careful design based on the incentives of the participants.  In common PoS designs, the signers of the new block receive an economical incentive (e.g., Ethereum). However, since in PoS creating forks is costless (as opposed to PoW), a participant may extend more than one of the existing forks (note that in PoW this does not make sense, as it means dividing your computing resources) to make sure it will get the block reward regardless of the winning chain. Ultimately, this leads to an endless number of forks and the blockchain not reaching consensus. What's worse, this may occur because of the monetary incentive, not due to a malicious attacker.

On the contrary, this does not hold in IPchain address since participants do not receive any incentive. The incentive is, as stated before, achieving a consistent view of the IP address space and having a secure Internet.

\subsection{Range Attacks}\label{security:shortr}
A range attack \cite{ethRangAtt} is performed by creating a fork some blocks back from the tip of the chain (it can be a few blocks $\sim$500 (short range attack) or a large amount $\sim$10000 (long range attack)). In this scenario, the attacker has privately fabricated a chain which (according to the consensus algorithm rules) will be selected over the original one. Benefits of this attack include gaining more stake on the blockchain (this attack could be part of a stake grinding attack) or rewriting the transaction history to erase a payment made in the original blockchain.

The simplest solution to this attack is adding a revert limit to the blockchain, forbidding forks going back more than N blocks.  This provides a means to solidify the blockchain.  However, nodes that have been offline for more than N blocks will need an external source that indicates the correct chain.  It has been proposed to do this out of band.  This is why some PoS algorithms are not purely trustless and require a small amount of trust.

\subsection{Stake Grinding}
Stake grinding refers to the manipulation of the consensus algorithm in order to progressively obtain more stake, with the goal of signing blocks more frequently with the ultimate goal of taking control of the blockchain.  It proceeds as follows: when the attacker has to sign a block, it computes all the possible blocks (varying the data inside them) to find a combination that gives the highest possibility of signing another block in the future.  It then signs this block and sends it to the network.  This procedure is repeated for all the next signing opportunities.  Over time, the attacker will sign more and more blocks until the consensus algorithm will always select the attacker to sign all blocks, thereby having taken control of the blockchain.

To prevent this attack, the source of randomness used to select the signers has to be hard to alter or to predict.
\subsection{Monopolies}
A crucial problem in PoS algorithms are monopolies, i.e. a single party controlling enough IP addresses so it can sign a significant proportion of new blocks, thus being able to decide which information is written in the chain (a 51\% attack in Bitcoin). However, in our case this is of less concern, because parties do not have an incentive to alter normal chain operation. Even if a single party could control the chain, rewriting data would end up impacting its own infrastructure, because some of its customers would be unable to access other networks. In other words, the incentive is a secure Internet, and any ISP benefits from this.


\subsection{Lack of participation}
Participants in a PoS algorithm will not always sign a block, since they might be offline when they are selected or lack incentives. Because of this, the final fraction of high-stakers that sign blocks can be very different from the full set of high-stakers.  The direct consequence of this situation is that the portion of participants that decide what goes into the blockchain can be a small set of
nodes.  If this participation is low enough, it can leave the control of the blockchain to a small amount of people/oligarchy, thus rising security concerns.
\subsection{Infrastructure Attacks}
This section presents attacks directed towards the underlying P2P network used to exchange information among the participants of the blockchain.
\subsubsection{DDoS Attacks}
Since blockchains are inherently based on P2P architectures, they present a higher degree of resistance to DDoS attacks than centralized server architectures, provided that the network has a significant number of participants.  In addition, it is always possible to keep an offline copy of the blockchain. 
\subsubsection{Transaction flooding}
A special type of DDoS attack consists in creating a large amount of legit transactions that transfer a small amount of tokens (i.e. delegate a lot of small IP prefixes).  If the number of transactions is large enough, the addition of new transactions can be significantly delayed because not all of them fit into a single block.  The effectiveness of the attack also depends on the throughput of the blockchain (transactions/second).  Simple solutions may be to limit the granularity upon which IP addresses can be split. Of course, only the legitimate holder of a large amount of IP address can perform this attack.
\subsubsection{Routing Attacks}   
The underlying P2P network in blockchains does not typically use any security mechanism, e.g. node authentication or integrity of network protocol messages.  This enables potentially disruptive attacks.  For example, specially located rogue nodes could drop new transactions, which would block updates on the blockchain and leave legit nodes isolated.  The effectiveness of this kind of attacks depends on how the P2P algorithm selects peers and the topology of the P2P network.

However, the most potentially dangerous attack of this type are network partitions, i.e. isolating a group of nodes from the rest of the network so they cannot communicate each other \cite{apostolaki2017}. The consequence of this attack is that two versions of the blockchain are created, one at each network partition.  When the partition disappears and the nodes reconnect one
of the two chains will be discarded, causing a service disruption. It is worth noting that Bitcoin has suffered similar attacks \cite{Greenberg2014}.
\subsubsection{Transaction Censorship}   
When a node adds a block it chooses arbitrarily which transactions are added into it, i.e. no specific rules control how transactions are added to a block.  This enables a node to selectively add some
transactions and intentionally exclude others, with the consequence that some transactions may be never added to the blockchain.  In the context of IP addresses, this may be performed by a competing ISP to prevent another ISP from executing a certain modification.  Possible solutions revolve around: giving more priority to older transactions (similarly to Bitcoin), or punishing nodes that exhibit this kind of behavior, e.g. removing part of their block of IP addresses or lowering their chance of adding blocks.

\end{document}